\documentclass[twocolumn,showpacs,preprintnumbers,amsmath,amssymb,APSl,prd,nofootinbib]{revtex4-1}

\usepackage{dcolumn}% Align table columns on decimal point
\usepackage{bm}% bold math
\usepackage{ifpdf}
\usepackage{hyperref}
\usepackage{dcolumn}
\usepackage{bm}
\usepackage[spanish,english]{babel}
\usepackage{amsfonts}
\usepackage{amssymb}
\usepackage{graphicx}
\usepackage[latin1]{inputenc}

\newcommand \be{\begin{equation}}
\newcommand \en{\end{equation}}
\newcommand \bea{\begin{eqnarray}}
\newcommand \ena{\end{eqnarray}}

\begin{document}

\title{Gauss-Bonnet black holes supported by a nonlinear electromagnetic field}

\author{D. Rubiera-Garcia} \email{drubiera@fudan.edu.cn}
\affiliation{Center for Field Theory and Particle Physics and Department of Physics, Fudan University, 220 Handan Road, 200433 Shanghai, China}

\date{\today}

\pacs{04.40.Nr, 04.50.-h, 04.50.Kd, 04.70.Bw}

\begin{abstract}

We study $D$-dimensional charged static spherically symmetric black hole solutions in Gauss-Bonnet theory coupled to nonlinear electrodynamics defined as arbitrary functions of the field invariant and constrained by several physical conditions. These solutions are characterized in terms of the mass parameter $m$, the electromagnetic energy $\varepsilon$ and the Gauss-Bonnet parameter $l_{\alpha}^2$. We find that a general feature of these solutions is that the metric behaves in a different way in $D=5$ and $D>5$ space-time dimensions. Moreover, such solutions split into two classes, according to whether they are defined everywhere or show branch singularities, depending on ($m, \varepsilon, l_{\alpha}^2$). We describe qualitatively the structures comprised within this scenario, which largely extends the results obtained in the literature for several particular families of nonlinear electrodynamics. An explicit new example, illustrative of our results, is introduced. Finally we allow non-vanishing values of the cosmological constant length $l_{\Lambda}^2$, and study the existence of new structures, in both asymptotically Anti-de Sitter and de Sitter spaces.

\end{abstract}

\maketitle

\section{Introduction} \label{sec:I}

The consideration of extensions of General Relativity (GR) containing higher-order powers in the curvature invariants is motivated by the fact that they arise in the quantization of fields in curved space-time \cite{quantumHC} and typically appear in several approaches to quantum gravity such as those based on string theory \cite{stringHC}. Indeed, it is expected that the effective Lagrangian of quantum gravity resulting from the expansion to the low-energy regime will generically include these higher-order curvature terms \cite{Cembranos}. In this sense, the finding and characterization of solutions, in particular, black holes, is of great interest to shed light on the understanding of this kind of modified, classical gravity theories and hopefully provide some insights on the low-energy limit of quantum gravity. However, it turns out that the class of gravitational actions made up as functions of curvature invariants such as $g_{\mu\nu}R^{\mu\nu}, R_{\mu\nu}R^{\mu\nu}$ and $ R_{\mu\nu\alpha\beta}R^{\mu\nu\alpha\beta}$ generically give rise to fourth-order differential field equations of motion and bring in ghosts. Nonetheless, as found by Lovelock in the 70's \cite{Lovelock} (see \cite{Lov-review} for reviews on this topic), there is a particular combination of such invariants in which the field equations contain only up to second-order derivatives of the metric and the quantization of the linearized theory is free of ghost \cite{Boulware85,Zumino86}. The Lovelock Lagrangian consists of a sum of dimensionally extended Euler densities as

\be
L_{LOV}=\sum_{k=0}^{[(D-1)/2]}c_kL_k, \label{lovelock}
\en
where [z] is the integral part of the number z, $D$ is the number of space-time dimensions, $c_k$ is the $k$-th order Lovelock parameter and $L_k$ are given by

\be
L_k=\frac{1}{2^k}\delta_{\mu_1 \nu_1 \cdot \cdot \cdot \mu_k \nu_k}^{\rho_1 \sigma_1 \cdot \cdot \cdot \rho_k \sigma_k}
R_{\rho_1 \sigma_1}^{\mu_1 \nu_1} \cdot \cdot \cdot R_{\rho_k \sigma_k}^{\mu_k \nu_k},
\en
where $\delta_{\mu_1 \nu_1 \cdot \cdot \cdot \mu_k \nu_k}^{\rho_1 \sigma_1 \cdot \cdot \cdot \rho_k \sigma_k}$ is the generalized totally antisymmetric Kronecker delta. For a given $D$ only terms with $k<(D-1)/2$ contribute to the equations of motion, while terms with $k>(D-1)/2$ do not and the case $k=(D-1)/2$ becomes a topological term. The zeroth and first order terms in the Lagrangian (\ref{lovelock}) corresponds to the cosmological constant and the Einstein-Hilbert action, respectively, while the second order

\be
L_{GB}=R_{\mu\nu\rho\sigma}R^{\mu\nu\rho\sigma}-4R_{\mu\nu}R^{\mu\nu}+R^2,
\en
is the Gauss-Bonnet (GB) Lagrangian. This is the simplest nontrivial modification of GR providing modified dynamics and second-order field equations as long as $D>4$. For $D>6$ one may consider the next order in the curvature invariants in (\ref{lovelock}) to provide additional modified dynamics, and so on. Finding exact, nontrivial solutions to the field equations of the family of Lagrangians (\ref{lovelock}) is, in general, a hard task, only achieved in some particular cases \cite{lovelock-sol}. Remarkably, the Lovelock family of Lagrangians lies at the crossroad of the metric and Palatini formulations of modified gravity, in the sense that the field equations in both formalisms turn out to be the same, as opposed to the general case \cite{Borunda}. This fact makes Lovelock gravities physically appealing and further supports the interest on them. Other approaches to higher-order gravity theories with second-order field equations, dubbed quasi-topological gravities, have been recently considered in the literature \cite{Quasi-topological}.

Both in GR, Gauss-Bonnet, and Lovelock gravities much interest has been paid to charged, nonrotating, black holes. This is due, on the one hand, to the fact that in this case exact solutions are more easily accessible and, on the other hand, that classical models have been attracted a great deal of attention in order to model the behaviour of charged particles. Historically, the first example of such models was the Born-Infeld (BI) Lagrangian, introduced in the thirties to remove the divergence of the electron's self-energy in classical electrodynamics \cite{BI}. Indeed, BI-like Lagrangians have been found to arise in the low-energy limit of string and D-Brane physics \cite{ST-BI}, which has raised a renewed interest on the consideration of nonlinear electrodynamics (NED) in a variety of gravitational backgrounds. For instance, electrostatic spherically symmetric black hole solutions of BI theory have been studied in the context of GR \cite{BI-Einstein}, in Anti-de-Sitter spaces in several dimensions \cite{BI-AdS}, in GB theory \cite{BI-Lovelock} and in $f(R)$ models \cite{or11}. Other physical motivations to consider NED models include the effective Lagrangian description of quantum electrodynamics effects \cite{EH}, the finding of regular black hole solutions in GR \cite{Regular,Regular2} or the existence of models for which Maxwell conformal invariance holds in any dimension \cite{Einstein-NED5}. When any of these modifications of Maxwell electrodynamics is coupled to a particular gravity theory, a different structure of the corresponding black hole solutions arise, which manifests in the structure of their singularities, or in the number and type of black hole horizons, both qualitatively and quantitatively \cite{Einstein-NED1,Einstein-NED2,Einstein-NED3, Einstein-NED4, Lovelock-HI,Lovelock-NED,Anninos, dilaton}.

The main result of this paper is to show that, no matter the physical motivation underlying a particular NED model, it is possible to establish general statements on the features of electro-static spherically symmetric (ESS) solutions in GB theory coupled to NED models \emph{without explicitly specifying the form of the Lagrangian density function}, provided that a number of physical reasonable conditions on the form of the NED Lagrangian density function (\emph{``admissibility"}) are assumed. This means that NED models are defined here as \emph{arbitrary} functions of the field invariants. These models fall into classes, according to the qualitative features of the corresponding black hole solutions, the same for all models within a given class. This framework largely broadens the class of NED models studied in the literature since, as opposed to the aforementioned studies, a full classification of the corresponding gravitating structures in any such admissible theories coupled to GB gravity takes place in terms of a few parameters. No explicit form for the NED function is needed. More specifically, these parameters are the relation between the flat-space ESS energy, $\varepsilon$ \footnote{That this energy be finite or not depends on the behaviour of the ESS field around the center of the solutions which, in turn, is just a consequence of the functional form of the matter Lagrangian density, see Sec.\ref{sec:III}.}, and the mass parameter, $m$, the number of space-time dimensions, $D$, and the value of the GB coupling constant, $l_{\alpha}^2$ (and the cosmological constant length, $l_{\Lambda}^2$, in asymptotically (Anti-)de Sitter backgrounds). In some cases the behaviour of these solutions, when defined everywhere, differs largely between the space-time dimensions $D>5$ and $D=5$.

As an illustrative example of our analysis, the third-order Lagrangian of quantum electrodynamics (the second-order is the well known Euler-Heisenberg Lagrangian of Quantum Electrodynamics \cite{EH}) is briefly discussed. In this work we only consider horizons with the usual spherical topology $k=1$.

\section{Integration of the field equations} \label{sec:II}

The $D$-dimensional action for the GB theory coupled to NED models reads

\be
S=\frac{1}{2\kappa^2} \int d^{D}x \sqrt{-g}\Big[(R-2\Lambda)+ \alpha L_{GB}]+ S_{NED}, \label{action}
\en
where $k^2=8\pi G_{D}$, with $G_{D}$ the $D$-dimensional gravitational constant, $\Lambda$ is the cosmological constant, and $\alpha$ is the GB constant with dimensions of (length)$^2$. As already mentioned, the GB term is a topological invariant in $D=4$ and thus it does not contribute to the field equations, so in order to obtain modified dynamics as compared to GR, in what follows $D \geq 5$ is assumed. The matter (NED) action is given by

\be \label{eq:NEDaction}
S_{NED}= \int d^{D}x \sqrt{-g} \varphi(X),
\en
where $\varphi(X)$ is an arbitrary function of the field invariant $X=-\frac{1}{2}F_{\mu\nu}F^{\mu\nu}$, constructed with the field strength tensor $F_{\mu\nu}=\partial_{\mu} A_{\nu}-\partial_{\nu}A_{\mu}$. In $D=4$ another field invariant can be constructed, namely, $-\frac{1}{2}F_{\mu\nu}F^{*\mu\nu}$, where $F^{*\mu\nu}=\frac{1}{2} \epsilon^{\mu\nu\alpha\beta}F_{\alpha\beta}$ is the dual of the field strength tensor. However, due to the dependence on $F^{*\mu\nu}$ such an invariant cannot be defined for $D>4$. In addition, for $D$ odd, one might add a Chern-Simons term to the action (\ref{action}), but we shall not consider it here and restrict ourselves to the electromagnetic field invariant $X$. Examples of nonlinear actions for the electromagnetic field coupled to gravity and considered in the literature include, among many others, Born-Infeld \cite{BI,BI-Einstein,BI-AdS,BI-Lovelock}, generalized Born-Infeld \cite{Einstein-NED1}, logarithmic \cite{Einstein-NED2}, Euler-Heisenberg \cite{EH,Einstein-NED3}, derivative corrections to Maxwell \cite{Einstein-NED4}, power-like and conformally-invariant \cite{Einstein-NED5}, Hoffman-Infeld \cite{Lovelock-HI}, Coulomb-like \cite{Coulomb-like}, or models leading to regular solutions \cite{Regular2}.

For a given (unspecified) $\varphi(X)$ function, and for ESS solutions, whose unique non-vanishing component is $E(r)=F^{tr}$, the components of the energy momentum tensor

\begin{equation}
{T_\mu}^{\nu}=2\varphi_X {F^\alpha}_{\mu} {F_{\alpha}}^{\nu} - {\delta_\mu}^{\nu} \varphi(X),
\end{equation}
(where $\varphi_X \equiv \frac{\partial \varphi}{\partial X}$) are obtained as

\begin{eqnarray} \label{eq:emcomponents}
T_t^t&=&T_r^r=2\varphi_X \vec{E}^2-\varphi(X) \\ T_{\theta_i}^{\theta_i}&=&T_{\theta_j}^{\theta_j}=-\varphi(X), i,j=2 \cdot \cdot \cdot D-1. \nonumber
\end{eqnarray}
On the other hand, the line element for a static, spherically symmetric space-time may be written as

\be \label{eq:line-element}
ds^2=e^{\nu(r)}dt^2-e^{\lambda(r)}dr^2-r^2d\Omega_{D-2}^2,
\en
where $d\Omega_{D-2}^2=d\theta_1^2+\sum_{i=2}^{D-2}\prod_{j=1}^{i-2} \sin^2 \theta_{j} d\theta_{i}^2$ is the metric on the unit ($D-2$) sphere. Due to the source symmetry $T_t^t=T_r^r$ it can be shown that the ($t,t$) and ($r,r$) components of the Einstein equations lead to a single independent metric function that may be written, without loss of generality, as $g_{\alpha}(r)=e^{\nu(r)}=e^{-\lambda(r)}$. Indeed, since the general procedure to obtain the metric function $g_{\alpha}(r)$ for a given NED source $T_t^t$ in GB theory has been nicely described in Ref.\cite{miskovic10} (and employed in several particular cases, see \cite{BI-Lovelock,Lovelock-NED,Anninos,Lovelock-HI}), we shall not repeat the derivation and instead briefly summarize the main steps. From the action (\ref{action}) the variational principle leads to the field equations

\begin{equation}
G_{\mu\nu}+\Lambda g_{\mu\nu}=R_{\mu\nu}-\frac{1}{2}g_{\mu\nu}R+\alpha G_{\mu\nu}^{GB}+\Lambda g_{\mu\nu}= \kappa^2 T_{\mu\nu},
\end{equation}
where the correction $G_{\mu\nu}^{GB}$ is given by

\begin{eqnarray}
G_{\mu\nu}^{GB}&=&2[RR_{\mu\nu}-2R_{\mu\alpha}R^{\alpha}_{\nu}-2R^{\alpha \beta}R_{\mu\alpha\nu\beta}+\nonumber \\ &+& R_{\mu}^{\alpha\beta\gamma}R_{\nu\alpha\beta\gamma}]- \frac{1}{2}g_{\mu\nu}L_{GB}.
\end{eqnarray}
For the line element (\ref{eq:line-element}) these equations lead to the relation ($\widetilde{\alpha}\equiv(D-3)(D-4) \alpha$)
\be
g_{\alpha}(r)-g_0(r)=\frac{\widetilde{\alpha}}{r^2}(1-g_{\alpha}(r))^2, \label{rel}
\en
 between $g_{\alpha}(r)$ and the solution with $\alpha=0$, which corresponds to the metric function that one would obtain for the very same problem in GR formulated in $D$ dimensions, i.e. static spherically symmetric solutions of the Einstein-Hilbert Lagrangian coupled to NED matter (\ref{eq:NEDaction}) and with a cosmological constant term. The finding of such solution $g_0(r)$ is obtained by taking advantage of the fact that $X=E^2(r,q)$ does not depend explicitly on the metric as a consequence of $g_{tt}g_{rr}=-1$. Using this, the field equations
$\nabla_{\mu}(\varphi_X F^{\mu\nu})=0$, for ESS fields, admit a first integral given by

\be \label{FI}
r^{D-2}\varphi_X E(r)=q,
\en
where $q$ is an integration constant, related to the physical charge $Q$ of the ESS field as

\be
Q^2= \frac{(D-2)(D-3)}{2} q^2.
\en
It is worth mentioning that (\ref{FI}) takes the same form as in absence of gravity. Moreover, it remains also unmodified for the GB theory, as a consequence of the source symmetry $T_t^t=T_r^r$.  Such a first integral determines the ESS field once the Lagrangian density function is given, and this field takes the same form, in the Schwarzschild-like coordinate system (\ref{eq:line-element}), as in the absence of gravitation in spherical coordinates. The Einstein equations in this case can be easily integrated using (\ref{FI}), leading to (in what follows we redefine $\tilde{\kappa}^2=\kappa^2 / \omega_{D-2}$, where $\omega_{D-2}=\frac{2\pi^{\frac{D-1}{2}}}{\Gamma((D-1)/2)}$ is the surface volume of the $(D-2)$-dimensional unit sphere and, subsequently, drop the tilde by notational simplicity)

\begin{equation}
g_0(r)=1-\frac{m}{r^{D-3}}+\frac{2\kappa^2}{(D-2)r^{D-3}} \varepsilon_{ex}(r,q)+\frac{r^2}{l_{\Lambda}^2}\label{metric0}
\end{equation}
where $m$ is an integration constant related to the ADM mass, $M$, of the solution as \cite{abbott82}

\begin{equation}
M=\frac{(D-2)\omega_{D-2}}{16\pi} m
\end{equation}
where we have defined

\begin{equation}
\varepsilon_{ex}(r,q)=\omega_{D-2}\int_r^{\infty}R^{D-2}T_0^0(R,q)dR \label{ext}
\end{equation}
which is physically interpreted as the energy of the ESS field outside of the sphere of radius $r$ in absence of gravity. In (\ref{metric0}) the (Anti-de) Sitter (AdS) radius $l_{\Lambda}^2=-\frac{(D-1)(D-2)}{2\Lambda}$ parameterizes the cosmological constant term. Once the solution in the Einstein gravity (\ref{metric0}) is known, the above equation (\ref{rel}) can be easily solved as

\be
g_{\alpha}(r)=1+\frac{r^2}{l_{\alpha}^2}\Big(1+\epsilon \sqrt{1+\frac{2l_{\alpha}^2}{r^{D-1}}\Big(m-\frac{2\kappa^2}{D-2}\varepsilon_{ex}(r,q)-
\frac{r^{D-1}}{l_{\Lambda}^2} \Big)}\Big),  \label{metric}
\en
where we have defined $l_{\alpha}^2=2\widetilde{\alpha}$. In the limit $\alpha \rightarrow 0$ this expression reduces to the one of (\ref{metric0}) and thus we recover the solution of the Einstein-NED-$\Lambda$ system, while in the limit $q \rightarrow 0$ we obtain the solution of Boulware and Deser \cite{Boulware85}. Note that there are two different branches for the solution (\ref{metric}) depending on $\epsilon=\pm 1$, which comes from taking a square root in the resolution of a quadratic equation for $g_{\alpha}(r)$. In the limit $\alpha \rightarrow 0$, the ``plus" branch leads to $g_{\alpha}^{+} \simeq 1+\frac{r^2}{l_{\Lambda}^2} - \frac{1}{r^{D-3}}(m-\frac{2\kappa^2}{(D-2)} \varepsilon_{ex}(r,q))$, which possesses an opposite sign for the gravitational mass. It has been argued by Boulware and Deser that in the vacuum case this branch is intrinsically unstable and the associated graviton becomes a ghost \cite{Boulware85}, suggesting that this branch is physically of less interest (see, however, \cite{deser03}). On the other hand, the ``negative" branch recovers the right GR limit, and thus in this work only this branch will be considered.

\section{The models and the ESS fields} \label{sec:III}

Up to now the discussion is valid for any NED model and static spherically symmetric electrovacuum solutions. Let us now specify the class of NED models that shall be considered throughout this paper. This analysis extends the one performed in \cite{dr10} to the $D$-dimensional case as detailed next. First we restrict ourselves to models for which the \emph{definiteness, derivability and single-valuedness} conditions on the NED function $\varphi(X)$ hold in all the domain of definition of $X>0$ covered by their ESS solutions (see Ref. \cite{dr09}). Such a condition is imposed in order to avoid multi-branched Lagrangian densities \cite{Bronnikov}, which carry potential singularities at the junction points in the effective geometry as seen by photons (see the analysis of \cite{Novello} in the case of GR).

The second requirement concerns the fulfillment of the positive definiteness of the energy functional in absence of gravity which, for a diagonal energy-momentum tensor, $T_{\mu\nu}=diag(\rho,p_r,p_2, \cdot \cdot \cdot, p_{D-1})$, reads simply $\rho=p_r>0$. Note that the weak energy condition (WEC) $T_{\mu\nu} \xi^{\mu} \xi^{\nu} \geq 0$, with $\xi^{\mu}$ a timelike vector, implies in addition $p_2= \cdot \cdot \cdot = p_{D-1}$, which holds automatically for the NED energy-momentum tensor (\ref{eq:emcomponents}), implying that our models will thus satisfy the WEC. Explicitly, this constraint on the energy density implies

\be
\rho=T_t^t=2X\varphi_X - \varphi(X) \geq 0, \label{pos}
\en
which must be satisfied everywhere for any field configuration. Though wormhole solutions violating the energy conditions (phantom energy models) have been considered in the literature \footnote{In these models it is usually assumed that violations of the energy conditions can occur due to quantum fluctuations, at least at some scales.} giving rise to a very active field of research \cite{Wormholes}, here we shall not get into such considerations and restrict ourselves to models satisfying the two conditions above.

To extract information from (\ref{pos}) we first assume that the energy density vanishes in vacuum ($X \equiv E^2=0$), which implies that $\varphi(0)=0$. By analyzing the inequality (\ref{pos}) one easily obtains that three conditions on the function $\varphi(X)$ must hold. From the fact that the ESS field can grow to arbitrarily large values, we must have $\varphi_X>0 (\forall X \neq 0)$, implying that $\varphi(X)$ is a strictly monotonically increasing function (excepting at $X=0$, where its derivative can vanish). If we assume the Lagrangian density to be defined everywhere, then the condition $\varphi(X)<0$ must hold in the region $X<0$. Finally, in the region $X\geq 0$ the inequality (\ref{pos}) implies that the function $\varphi(X)/\sqrt{X}$ must be positive increasing. If we add to this discussion the first-integral (\ref{FI}) it follows immediately that these \emph{admissibility} conditions endorse the everywhere definiteness, strictly monotonic behaviour, and single-branched character of the (asymptotically vanishing) ESS solutions. An important consequence of these admissibility conditions is the fact that the function $\varepsilon_{ex}(r,q)$ in (\ref{ext}) becomes a monotonically decreasing and concave function of $r$, as can be easily seen by double derivation of (\ref{ext}) and taking into account the admissibility conditions established above.

Let us now study the behaviour of the total energy in the ESS field in $D$ dimensions, which is obtained as

\bea
\varepsilon(q)&=& \omega_{D-2} \int_0^{\infty}r^{D-2}T_t^t(r,q)dr \nonumber \\
&=& \omega_{D-2} \int_0^{\infty} dR(2qE-R^{D-2} \varphi) \\
&=& q^{\frac{D-1}{D-2}} \varepsilon(q=1), \nonumber \label{energy}
\ena
where $\varepsilon(q=1)$ is the solution of unit charge. The conditions for the finiteness of (\ref{energy}) are easily obtained using Eq.(\ref{FI}), and amounts to a simple extension of the results derived in Ref.\cite{dr10}, for the case of arbitrary $D \geq 4$.

In order for (\ref{energy}) to converge at $r \rightarrow \infty$, we assume that the field vanishes asymptotically as (see the definition of $T_t^t$ in Eq.(\ref{eq:emcomponents}))

\be
E(r) \sim \frac{\beta}{r^p}, \label{asymp}
\en
where $\beta$ is a constant, and we must determine next the value of the parameter $p>$ for convergence of the energy. Using (\ref{FI}) the associated behaviour for the Lagrangian density function is given by

\be \label{eq:A1}
\varphi(X) \sim X^{\alpha},
\en
where $\alpha=\frac{p+D-2}{2p}$. Inserting these behaviours in (\ref{energy}) and noting that $r^{D-2} \varphi \sim E$ it follows that finiteness of the energy implies $p>1$. Therefore, one has $\alpha<(D-1)/2$, with a lower bound given by $\alpha>1/2$ (corresponding to $p \rightarrow \infty$), Thus the ESS field approaches its asymptotically vanishing values slower ($p<D-2$), equal to ($p=D-2$) or faster ($p>D-2$) than the $D$-dimensional Coulombian one. Consequently, a NED model for which the energy of the ESS field at $r \rightarrow \infty$ in four space-time dimensions is finite, leads also to asymptotically finite-energy ESS solutions in the $D$-dimensional case ($D>4$).

For the central region $r \rightarrow 0$ there are, for admissible models, two field behaviours compatible with the finite-energy requirement. In class A1 the ESS field behaves as in (\ref{asymp}) (so it diverges at $r \rightarrow 0$) but now with $\alpha>(D-1)/2$ for convergence of the energy there, which implies $0<p<1$. The Lagrangian density behaves as in Eq.(\ref{eq:A1}) with the same relation between $\alpha$, $p$ and $D$. This implies that if we have a model whose finite-energy ESS solutions belongs to this class in $D=4$ dimensions (implying $\alpha>3/2$), they become of divergent-energy for some larger $D$. An explicit example of this is the EH model \cite{EH}, defined by a Lagrangian density function (in $D=4$) $\varphi(X,Y)=\frac{X}{2}+\xi\left(4X^2+7Y^2\right),\xi>0$. It can readily checked that this model contains finite-energy ESS solutions in $D=4$ dimensions, but not for $D>4$. However higher-dimensional finite-energy ESS solutions of this class A1 can be obtained by supplementing the EH Lagrangian with higher powers of $X$ (see Sec.\ref{sec:IV-III}).

It is worth pointing out that if the A1 family is allowed to cover the case with $p>1$, this leads to another family of ESS solutions (A0 class), now with divergent energy. The $D$ dimensional Coulomb field ($p=D-2$) belongs to this family. Given the monotonically decreasing and concave character of $\varepsilon_{ex}(r,q)$ for admissible models, the behaviour of all the models in this family corresponds to taking $\varepsilon \rightarrow \infty$ in the discussion of Sec.\ref{sec:IV} below and, as a consequence, its behaviour will be similar to the Reissner-Nordstr\"om-GB one (first studied in the first of Refs.\cite{BI-Lovelock}).

On the other hand, in the class A2 the ESS field attains a finite value at the center, behaving there as

\be
E(r) \sim a - br^{\sigma}, \sigma>0,
\en
while the behaviour of the Lagrangian density becomes, by using (\ref{FI}), in

\be
\varphi(X)\sim \left(\sqrt{X}-a\right)^{\gamma}+\Delta, \label{laga2}
\en
where $\Delta$ is an integration constant and the restriction $\gamma=\frac{\sigma-D+2}{\sigma}<1$ must hold for finiteness of the energy. In this family there are two different behaviours for the Lagrangian density functions. If $0<\gamma<1$ ($\sigma>D-2$) the Lagrangian density takes a finite value around $r\rightarrow 0$ given by the value $\Delta=\varphi(a^2)$, while for $\gamma < 0$ ($\sigma < D-2$) the Lagrangian diverges as $X\rightarrow a^2$. The case $\gamma=0$ ($\sigma= D-2$) is singular, behaving as

\be
\varphi(X)\sim -\ln\left(a-\sqrt{X}\right),
\en
but shows the same behaviour as the $\gamma<0$ one. The Born-Infeld Lagrangian \cite{BI} is a well known member of this family, with $\sigma=4$.

\begin{table}
 \begin{center}
   \begin{tabular}{| c | c | c |  c | }
        \hline
      Models & Field & Lagrangian & Finiteness of $\varepsilon$  \\ \hline
      $r \rightarrow \infty$& $E(r) \sim \beta/r^p$ & $\varphi(X) \sim X^{\alpha}$ & $\alpha < (D-1)/2$  \\ \hline
      A1 & $E(r) \sim \beta/r^p$ & $\varphi(X) \sim X^{\alpha}$ & $\alpha > (D-1)/2$ \\ \hline
      A2 & $E(r) \sim a-br^{\sigma}$ & $\varphi(X) \sim (a-\sqrt{X})^{\gamma}$ & $\gamma<1$   \\ \hline
   \end{tabular}
 \caption{The NED models and their ESS solutions, together with the condition for finiteness of $\varepsilon$ in Eq.(\ref{energy}). In this table $\alpha=(p+D-2)/(2p)$ and $\gamma=(\sigma-D+2)/\sigma$.}
  \label{table:I}
 \end{center}
\end{table}

\section{Gauss-Bonnet-NED black holes} \label{sec:IV}

Having discussed the behaviours of the ESS field compatible with the admissibility requirement and the finiteness of the energy, as well as the consequences for the qualitative behaviour of the function $\varepsilon_{ex}(r,q)$ in (\ref{ext}), we proceed to study their black hole solutions within GB theory. In this section we shall deal with asymptotically flat solutions so we set the cosmological constant term to zero ($\Lambda=0$). As we shall see, all the cases with $D>5$ posses the same structure of horizons, but the behaviour of the metric in some of the cases for $D=5$ shows differences, leading to some new structures. This is due to the fact that in the equation of the horizons (if any), i.e. the solutions of $g_{tt}=0$, for the different configurations

\be
\frac{2\kappa^2}{D-2}\varepsilon_{ex}(r_h,q)=m-\frac{l_{\alpha}^2}{2}r_h^{D-5}-r_h^{D-3}, \label{hor}
\en
the right-hand-side at $r_h \rightarrow 0$ takes the value $m$ for $D>5$ but $m-l_{\alpha}^2/2$ in the $D=5$ case. Now, by taking into account the monotonic and concave character of $\varepsilon_{ex}(r,q)$ for admissible models, the horizons can be obtained from the cut points between the curve $\frac{2\kappa^2}{D-2} \varepsilon_{ex}(r,q)$ and the beam of curves $m-r^{D-3}-(l_{\alpha}^2/2) r^{D-5}$, corresponding to different values of $m$ once $l_{\alpha}^2$ is fixed (see Fig.\ref{fig:2}). This leads to an immediate classification of the number of possible horizons. Indeed, since the large-$r_h$ behaviour of this beam is governed by the term $-r_h^{D-3}$, this implies that it can be at most two cut points between the curves in Eq.(\ref{hor}), regardless of the sign and value of $l_{\alpha}^2$. However, a new feature arises here as compared with the GR case, since the term

\begin{equation}
a(r)=1+\frac{2l_{\alpha}^2}{r^{D-1}}\left(m-\frac{2\kappa^2}{D-2}\varepsilon_{ex}(r,q)\right), \label{eq:a(r)}
\end{equation}
inside the square root appearing in (\ref{metric}) can become negative, leading to a complex metric for radius smaller than a certain $r_S$, which is the solution of the equation

\be
r_S^{D-1}+2l_{\alpha_0}^2 \left(m-\frac{2\kappa^2}{D-2} \varepsilon_{ex}(r_S,q)\right)=0, \label{sing}
\en
where $l_{\alpha_0}^2$ is the value for which $a'(r_S)=0$ is satisfied, corresponding to

\be
(D-1)(D-2)+4\kappa^2 l_{\alpha_0}^2T_t^t(r_S,q)=0, \label{bs}
\en
This is a new kind of singularity, dubbed branch singularity (BS), arising at a non-vanishing horizon radius. This fact splits the solutions into those defined everywhere and those exhibiting branch singularities, according to the existence or not of a value $l_{\alpha_0}^2$. On the other hand, the asymptotic behaviour of the metric compatible with the finiteness of the energy there (see Eqs.(\ref{asymp}) and (\ref{eq:A1})) is given by

\bea \label{eq:a-behaviour}
g_{\alpha}(r) &\sim& 1+\frac{r^2}{l_{\alpha}^2}\Big(1+\epsilon
\Big[1+\frac{2 l_{\alpha}^2}{r^{D-1}}\Big(\Big(m-\nonumber \\ &-&  \frac{4\kappa^2 (D-2) \beta q}{(p-1)(p+D-2)r^{p-1}} \Big) \Big]^{1/2}\Big),
\ena
so for asymptotically coulombian fields ($p=D-2$ in Eq.(\ref{asymp})) we recover the GB-Reissner-Nordstr\"om solution obtained by Wiltshire in Ref.\cite{BI-Lovelock}.

For all the GB-NED solutions the sign of the quantity

\begin{equation}
\Sigma(q)=m-\frac{2\kappa^2}{D-2}\varepsilon(q), \label{Sigma}
\end{equation}
as well as the one of $l_{\alpha}^2$, becomes essential for their proper characterization. Let us now characterize the Gauss-Bonnet-NED black holes by analyzing first those whose metric is defined everywhere.

\subsection{Solutions defined everywhere} \label{sec:IV-I}

\subsubsection{$\Sigma(q)\geq 0, l_{\alpha}^2>0$} \label{sub:1}

\begin{figure}
\begin{center}
\includegraphics[width=8cm,height=5cm]{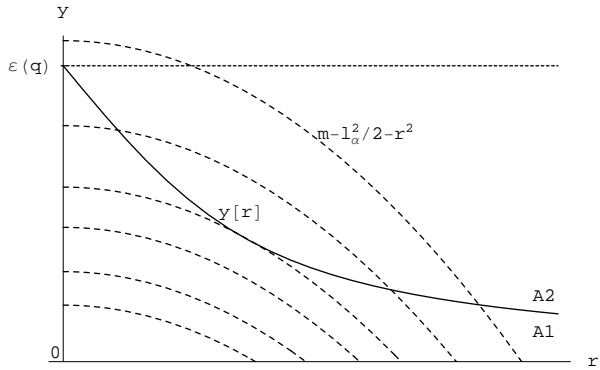}
\caption{\label{fig:epsart} Qualitative procedure to obtain the number of horizons in admissible models and solutions defined everywhere (see Sec.\ref{sec:IV-I}). The solid curves represent $y=\frac{2\kappa^2}{D-2}\varepsilon_{ex}(r,q)$ for case A2 (finite slope at $r \rightarrow 0$) and A1 (divergent derivative). Each curve begins at the value $\frac{2\kappa^2}{D-2}\varepsilon(q)$ (these curves have been normalized in this figure to some value $\varepsilon(q)$, by simplicity), see Eq.(\ref{hor}). The cut points with the beam of (dashed) curves $m-l_{\alpha}^2/2-r^2$ (with $l_{\alpha}^2>0$, and different values of $m$) give the horizons of the different configurations. Each curve of this beam takes the value $m$ when $D>5$ and  $m-l_{\alpha}^2/2$ for $D=5$. The case with $l_{\alpha}^2<0$ would show a positive slope at $r \rightarrow 0$ for the beam of curves, but the number of cut points with $y$  are the same as in the $l_{\alpha}^2>0$ case. For BS solutions only cut points corresponding to $r>r_S$ are relevant for the existence of horizons.} \label{fig:2}
%figure caption%
\end{center}
\end{figure}

As seen in Fig.\ref{fig:2}, the horizons (if any) of the different configurations are obtained though the cut points between the beam of curves $m-r_h^{D-3}-(l_{\alpha}^2/2)r_h^{D-5}$ and the (monotonically decreasing) curve $\frac{2\kappa^2}{D-2}\varepsilon_{ex}(r,q)$. From the positivity of $l_{\alpha}^2$ any curve of the beam in the right-hand-side of Eq.(\ref{hor}) for $D>5$ is a strictly monotonically decreasing function, starting from $m \geq \frac{2\kappa^2}{D-2}\varepsilon_{ex}(r,q)$, and thus there is a single cut point with $\varepsilon_{ex}(r,q)$ in this case. For $D>5$ the metric at the center behaves as

\be
g_{\alpha}(r\sim 0)\backsimeq 1-\frac{(2l_{\alpha}^2 \Sigma(q))^{1/2}}{l_{\alpha}^2 r^{\frac{D-5}{2}}} + \ldots , \label{met}
\en
and, consequently, it diverges there to $-\infty$, which, together with the asymptotic behaviour in (\ref{eq:a-behaviour}), confirms that there is a black hole solution with a single horizon $r_h$, given by the solution of the equation (\ref{hor}).

However, for $D=5$, at $r \rightarrow 0$ the above beam of curves takes the value $m-\frac{l_{\alpha}^2}{2}\gtrless \frac{2\kappa^2}{3}\varepsilon(q)$. When $>$ holds, there is a single cut point (see Fig.\ref{fig:2}) between the curve $\frac{2\kappa^2}{D-2}\varepsilon_{ex}(r,q)$ and the beam of curves in the right-hand-side of Eq.(\ref{hor}), while two, one (degenerate) or none appear when $<$ holds. The metric at the center takes a (finite and $<1$) value given by

\be
g_{\alpha}(r\rightarrow 0) \simeq 1-\frac{(2l_{\alpha}^2 \Sigma(q))^{1/2}}{l_{\alpha}^2} + \ldots .\label{finite-met}
\en
The derivative of the metric there depends on the central-field behaviour of the ESS fields. For the A1 class the leading behaviour term is given by ($0<p<1$)

\be
g'_{\alpha}(r\rightarrow 0) \simeq - \frac{4\kappa^2 q \alpha}{(p+3)(2l_{\alpha}^2 \Sigma(q))^{1/2}r^{p}} + \ldots ,
\en
(where $g' \equiv dg/dr$) which diverges to $-\infty$. In contrast, for the class A2 this derivative takes a finite (negative) value given by
\be
g'_{\alpha}(r \rightarrow 0) \simeq - \frac{4\kappa^2 q b}{3(2l_{\alpha}^2 \Sigma(q))^{1/2}} + \ldots .
\en
Consistently with the previous considerations, if $\Sigma(q)\geq l_{\alpha}^2/2$ then the metric around the center is negative and there is a black hole with a single horizon. Extreme black holes may arise (see Fig.\ref{fig:2}) in the case $ 0<\Sigma(q)< l_{\alpha}^2/2$ (now the metric at the center is negative). They are defined by the conditions $g_{\alpha}(r_h)=g_{\alpha}'(r_h)=0$, which lead to

\be
\frac{\kappa^2}{3}r_{hextr}^2(q)T_0^0(r_{hextr},q)=1. \label{eq:m-extre}
\en
Their mass is given by

\begin{equation}
m_{hextr}(q)= \frac{1}{4} \left[r_{hextr}^2 + l_{\alpha}^2  +2\kappa^2 q \Phi(r_{hextr},q) \right],
\end{equation}
where an integration by parts of the term $\varepsilon_{ex}(r,q)$ has been performed, and the quantity $\Phi(r_{hextr},q)=\int_{r_{hextr}}^{\infty}E(r,q)dr$ is the electric potential measured at infinity with respect to the horizon. For any ESS field of the form (\ref{asymp}) the potential at infinity vanishes and then we have $\Phi=A_0(r_{hextr},q)$.

In this $D=5$ case, besides extreme black holes ($m=m_{extr}(q)$), there may be black holes with inner and outer horizons ($m>m_{extr}(q)$), or naked singularities otherwise ($m<m_{extr}(q)$). Let us mention that for the critical value of the mass $m=\frac{2\kappa^2}{3}\varepsilon(q)$, which corresponds to a critical charge $q_{crit}=\left(\frac{3m}{2\kappa^2 \varepsilon(q=1)} \right)^{\frac{3}{4}}$ (so $\Sigma=0$, see Eqs.(\ref{energy}) and (\ref{Sigma})), the metric at the center is $g_{\alpha}(r=0)=1$. Such solutions are on the verge of becoming into those of subcases with branch singularities (\ref{sub:4}) and (\ref{sub:5}) (see subsection \ref{sec:IV-II}) when $m$ is decreased.

\subsubsection{$\Sigma< 0; l_{\alpha 0}^2<l_ {\alpha}^2<0$} \label{sub:2}

In this case the beam of curves in Eq.(\ref{hor}) can cut the curve $\varepsilon_{ex}(r,q)$ two, one (degenerate) or zero times, regardless of the value of $D$, since $m-l_{\alpha}^2/2-r_h^2 \mid_{r \rightarrow 0}$ is always below $\frac{2\kappa^2}{D-2}\varepsilon(q)$. For this class of solutions an extreme black hole is formed for the tangent cut point to the beam (\ref{hor}), leading to the extreme black hole radius $r_{extr}(q)$, obtained as

\bea
\kappa^2 r_{hextr}^4T_0^0(r_{hextr},q)&=&\frac{(D-2)(D-3)}{2}r_{hextr}^2+ \nonumber \\ &+& \frac{(D-2)(D-5)}{4}l_{\alpha}^2 ,\label{ebh}
\ena
of which Eq.(\ref{eq:m-extre}) is a particular case. As in subcase (\ref{sub:1}) with $D=5$, the associated extreme black hole mass in these solutions for any $D$ can be obtained by replacing the condition (\ref{ebh}) into Eq.(\ref{hor}), leading to

\be
m_{hextr}(q)=\frac{1}{D-1}\left[r_{hextr}^{D-3}+l_{\alpha}^2 r_{hextr}^{D-5} +2\kappa^2 \Phi(r_{hextr},q) \right]. \label{ebhm}
\en
However, although the structure of horizons is the same in the $D>5$ and $D=5$ cases (naked singularities, extreme black holes or two-horizon black holes) the behaviour of the metric at the center is not: as $r \rightarrow 0$ the metric is obtained as in Eq.(\ref{met}), so it diverges there to $+\infty$ when $D>5$, while in the case $D=5$ it takes a finite value given by Eq.(\ref{finite-met}) which, as opposed to the previous subcase, is now larger than one.

All these everywhere defined solutions have a curvature singularity at the center, due to the divergence of some of the curvature invariants such as $R, R_{\mu\nu}R^{\mu\nu},R_{\mu\nu\alpha\beta}R^{\mu\nu\alpha\beta}$. This is an unavoidable feature for the models studied here (classes A0, A1 and A2), resulting from the admissibility requirement. However, some models studied in the literature in the case of GR \cite{Regular2} have been found to support singularity-free gravitating solutions. In such a case the ESS field at the center vanishes and, as a consequence of Eq.(\ref{FI}), neither the ESS field nor $\varepsilon_{ex}(r,q)$ are no longer monotonically decreasing and the associated Lagrangian densities correspond to multivalued functions \cite{Bronnikov}. As a consequence the energy density may become non-positive definite. Let us also note that another Lagrangian with vanishing (at the center) ESS solutions is the Hoffman-Infeld model studied by Aiello et al \cite{Lovelock-NED}, given by a Lagrangian density of the form

\begin{equation}
L_{HI}=2\beta^2 (1-\eta(X)-\log(X)),
\end{equation}
where $\eta(X)=\frac{X}{2\beta^2} (1-\sqrt{1-X/\beta^2})^{-1}$. These solutions in GB gravity present a double-peak behaviour for the temperature function. Such classes of ESS solutions violate the admissibility conditions of section \ref{sec:III}, and shall not be considered here.

\subsection{Solutions with branch singularities} \label{sec:IV-II}

As already discussed, in some cases the equation of the horizons (\ref{hor}) is defined only beyond the singularity radius $r_S$ and thus in our procedure of obtaining the horizons the cut points between the curve $y=\frac{2\kappa^2}{D-2} \varepsilon_{ex}(r,q)$ and the beam of curves in (\ref{hor}) makes sense only for $r>r_S$. There are three cases to be analyzed separately:

\subsubsection{$\Sigma \geq 0;l_{\alpha}^2<0$} \label{sub:3}

At the BS point given by the Eq.(\ref{sing}) the metric in this case takes the value

\be
g(r_S)=1+\frac{r_S^2}{l_{\alpha}^2}<1, \label{met3}
\en
while the leading terms of its derivative there are given by

\be
g'_{\alpha}(r_S)\backsimeq - \frac{a'(r_S) r_S}{2(D-2)l_{\alpha}^2 \left(1+\frac{2l_{\alpha}^2}{r_S^{D-1}}
(m-\frac{2\kappa^2}{D-2}\varepsilon_{ex}(r_S,q))\right)^{1/2}} + \ldots , \label{der3}
\en
(see Eq.(\ref{eq:a(r)}) for the definition of $a(r_S)$). In the present case the derivative in (\ref{der3}) is positive as we approach $r-r_S \rightarrow 0^+$ since $a'(r_S)<0$. Now the number of cut points between $\frac{2\kappa^2}{D-2}\varepsilon_{ex}(r,q)$ and the beam of curves in (\ref{hor}) depends on the ratio $r_S^2/l_{\alpha}^2 \gtreqless -1$. Indeed, when $0>r_S^2/l_{\alpha}^2>-1$ the metric is everywhere positive and thus we are led to a naked singularity, while if $r_S^2/l_{\alpha}^2<-1$ we have a black hole with a single horizon. The location of such a horizon merges with the one of the BS point in the limit $r_S^2/l_{\alpha}^2 \rightarrow -1$ . Note, however, that the conditions determining the existence of BS (the values of $l_{\alpha_0}$ and $\Sigma(q)$) are not independent; indeed they are related by Eq.(\ref{bs}), since the NED energy-momentum tensor enters into both of them.

\subsubsection{$\Sigma< 0;l_{\alpha}^2>0$} \label{sub:4}

As in subcase (\ref{sub:3}) the metric at the BS point $r_S$ is finite, but now  $g(r_S)=1+r_S^2/l_{\alpha}^2>1$ and its slope there, given by Eq.(\ref{der3}), becomes negative in this case. Consequently the associated gravitational configurations correspond to black holes with two horizons, extreme black holes or naked singularities, depending on the value of the mass as compared with the extreme one, $m_{extr}$ given by Eq.(\ref{ebhm}).

\subsubsection{$\Sigma< 0; l_{\alpha}^2<l_ {\alpha 0}^2<0$} \label{sub:5}

In this case the BS condition (\ref{sing}) is satisfied twice, i.e., $r_S<r_{S'}$, and, consequently, the metric is only defined for $r<r_S$ and $r>r_{S'}$. Since $\varepsilon_{ex}(r,q)$ is a monotonically decreasing function, and larger than $m$ at $r \rightarrow 0$, as we increase the radius, the function $a(r)$ becomes negative in an interval (when $l_{\alpha}^2<l_{\alpha 0}^2$), before becoming positive again. The metric diverges at the center to $+\infty$ as a consequence of Eq.(\ref{met}). As pointed out in the first Ref.\cite{BI-Lovelock}, for an observer in the asymptotic region of space-time only the region beyond the outer singularity radius is physically accessible, which implies that in this case we are led to single-horizon black holes or naked singularities, depending on $g_{\alpha}(r_{S'})\lessgtr 0$.

Consequently we see that black hole solutions in GB theory coupled to NED models supporting finite-energy ESS fields [see Table \ref{table:II}] somewhat interpolate between GB-Reissner-Nordstr\"om-type solutions ($\varepsilon(q) \rightarrow \infty$) with a timelike singularity (subcases \ref{sub:2}, \ref{sub:4} and \ref{sub:5}) and GB-Schwarzschild-type solutions ($\varepsilon(q)=0$) with spacelike singularities (subcases \ref{sub:1} and \ref{sub:3}).

\begin{table}
 \begin{center}
   \begin{tabular}{| c | c | c |  c | }
        \hline
      Parameters & Range & Horizons $D>5$ & Horizons $D=5$  \\ \hline
      $\Sigma(q)\geq 0, l_{\alpha}^2>0$ & DE & 1 & 2, 1(e), 0, or 1  \\ \hline
      $\Sigma< 0; l_{\alpha 0}^2<l_ {\alpha}^2<0$ & DE & 2, 1(e), 0  & 2, 1(e), 0 \\ \hline
      $\Sigma \geq 0;l_{\alpha}^2<0$ & BS & 1, 0 & 1, 0   \\ \hline
      $\Sigma< 0;l_{\alpha}^2>0$ & BS & 2, 1(e), 0 & 2, 1(e), 0   \\ \hline
      $\Sigma< 0; l_{\alpha}^2<l_ {\alpha 0}^2<0$ & BS & 1, 0 & 1, 0    \\ \hline
   \end{tabular}
 \caption{The GB black holes for admissible models, and the corresponding number of horizons. In this table we use the labels: DE: defined everywhere, BS: branch singularities, e: extreme}
  \label{table:II}
 \end{center}
\end{table}

\subsection{A particular model} \label{sec:IV-III}

As an example of the above solutions, which captures the main features of the analysis performed here, let us consider a model belonging to the A1 class of solutions, given by the Lagrangian density

\be
\varphi(X)=\frac{X}{2}+\alpha_1 X^2+ \alpha_2 X^3, \label{laga1}
\en
with $\alpha_1$ and $\alpha_2$ being positive constants, in order to satisfy the admissibility conditions. In $D=4$ the two first terms in (\ref{laga1}) define the Euler-Heisenberg effective Lagrangian (in this case an additional term in $Y^2$ must be added to (\ref{laga1}), which vanishes for ESS solutions in $D=4$ and cannot be defined in $D>4$) of quantum electrodynamics (QED) \cite{EH}. The EH model has been shown to contain finite-energy ESS solutions in $D=4$ \cite{dr09}. However, for $D>4$ the term in $X^3$ must be added in order to keep the energy finite. For $D>6$ higher order terms in $X$ must be added to (\ref{laga1}) for this finite character of the energy to hold. Let us note that these terms containing higher powers in the field invariant $X$ arise from a low-energy expansion of QED, once the heavy degrees of freedom are integrated out in the path integral of the original action \cite{Dobado} (see also \cite{Pir}). By simplicity, here we shall restrict ourselves to the cases $D=5,6$, which will allow us to illustrate the different structures found here.

\begin{figure}
\begin{center}
\includegraphics[width=8cm,height=5cm]{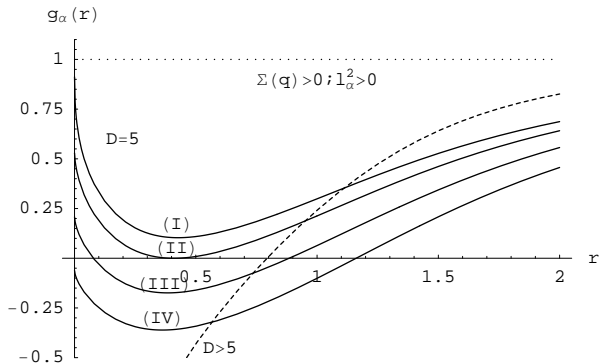}
\caption{\label{fig:epsart} Behaviour of the metric function for the model (\ref{laga1}) with the parameters
$\alpha_1=\alpha_2=q=1,l_{\alpha}^2=2$, in the case $\Sigma(q)>0, l_{\alpha}^2>0$. In $D=5$ the energy is $\varepsilon(q=1) \simeq 2.508$ and there are four structures: (I) $m=2.52$: naked singularity, (II) $m\simeq 2.841$: extreme black hole, (III) $m=3.45$: two-horizons black hole and (IV) $m=4.2$: single-horizon black hole. For $D=6$ (dashed curve, $m=4.25$) we find single-horizon black hole. All solutions are asymptotically flat.} \label{fig:3}
%figure caption%
\end{center}
\end{figure}

Using (\ref{FI}) the field behaviour at $r\rightarrow \infty$ and as $r \sim 0$ becomes

\be
E(r \rightarrow \infty,q) \sim \frac{2q}{r^{D-2}} \hspace{0.1cm};\hspace{0.1cm} E(r \sim 0,q) \sim \left(\frac{q}{3\alpha_2}\right)^{1/5}\frac{1}{r^{\frac{D-2}{5}}}, \label{field}
\en
and thus we are dealing with an asymptotically coulombian ESS field, belonging to the class A1 as $r \sim 0$ if $D<7$, as can be seen from (\ref{field}) and the considerations of section III. The behaviour of the associated energy density in those limits is given by

\bea
T_0^0(r\rightarrow \infty,q)&\sim& \frac{q^2}{2r^{2(D-2)}} \nonumber \\
T_0^0(r \sim 0, q) &\sim& 5\left(\frac{q}{3\alpha_2^{1/6}}\right)^{\frac{6}{5}}\frac{1}{r^{\frac{6(D-2)}{5}}}.
\ena
The total energy can be calculated by integrating by parts in (\ref{energy}) and using $y=E(r)$ as the integration variable, leading to (an additional term, vanishing for the finite-energy ESS solutions, has been omitted in this formula)

\be
\varepsilon(q)=\frac{4\omega_{D-2}}{D-1}q^{\frac{D-1}{D-2}}\int_{y(0)}^{y(\infty)} \frac{dy}{\left(y \cdot \varphi_X(X=y^2)\right)^{\frac{1}{D-2}}},
\en
which, for the family (\ref{laga1}) reads

\be
\varepsilon(q)=\frac{4\omega_{D-2}}{D-1}q^{\frac{D-1}{D-2}} \int_0^{\infty} \frac{dy}{\left(y(\frac{1}{2}+2\alpha_1 y^2+ 3\alpha_2 y^4)\right)^{\frac{1}{D-2}}},
\en
whose value can be obtained once the model parameters $\alpha_1,\alpha_2$ are given. It can be easily checked that this energy is always finite for $D<7$, regardless of the values of the constants $\alpha_1,\alpha_2$, as expected. For this model we failed to explicitly work $\varepsilon_{ex}(r,q)$ out but, however, it is possible to characterize numerically the different gravitational configurations for this model\footnote{By simplicity, in this example we take units in which $\omega_{D-2} \kappa^2=1$.}, using Eq.(\ref{metric}). In Fig.\ref{fig:3} we have plotted the behaviour of $g_{\alpha}(r)$ for the model parameters $\alpha_1=\alpha_2=1$, unit charge (for which $\varepsilon(D=5,q=1) \simeq 2.508$ and $\varepsilon(D=6,q=1) \simeq 3.446$), $l_{\alpha}^2=2$ and $\Sigma>0$. As expected, the structures for this case are those discussed in subsection \ref{sec:IV-I}: we see that in the $D=5$ case the metric around the center is finite and its slope diverges there to $-\infty$, leading to four classes of configurations (naked singularities, extreme black holes and black holes with two or a single non-degenerate horizon), while for $D=6$ the metric diverges to $-\infty$ as $r \sim 0$ and there is a black hole with a single horizon.

\begin{figure}
\begin{center}
\includegraphics[width=10cm,height=6cm]{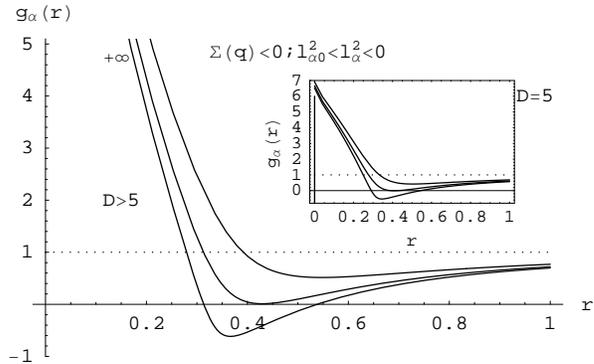}
\caption{\label{fig:epsart} Behaviour of the metric function for the model (\ref{laga1}) with the parameters $\alpha_1=\alpha_2=1$ and $q=1$ in the case
$\Sigma(q)<0, l_{\alpha0}^2<l_{\alpha}^2<0,l_{\alpha}^2=-0.05$ and for $D=6$. As in Fig.\ref{fig:3} the set of solid lines (diverging to $+\infty$ as $r \rightarrow 0$) corresponds to different values of $m$, leading to naked singularities,
extreme black holes or two-horizon black holes. The plot in the small frame shows the behaviour for $D=5$. In this case the metric as $r \rightarrow 0$ is finite but the associated structures are the same as for $D=6$.} \label{fig:4}
%figure caption%
\end{center}
\end{figure}

In Fig.\ref{fig:4} we have plotted the metric behaviour for the second class of solutions defined everywhere (see subsection \ref{sec:IV-I}), corresponding to $\Sigma(q)<0$ and $l_{\alpha 0}^2<l_{\alpha}^2=-0.05<0$, with model parameters $\alpha_1=\alpha_2=1$ and a unit charge. For $D=6$ the metric at the center diverges to $+\infty$ and one finds naked singularities, extreme black holes or two-horizon black holes. Similar structures are found for $D=5$ (see the small plot in Fig.\ref{fig:4}) but in this case the metric at the center is finite and larger than one. The BS solutions of subsection \ref{sec:IV-II} for this model can be obtained in a similar way. All these results for this particular model are in complete agreement with the general statements of the previous sections.

\section{Non-vanishing cosmological constant $\Lambda \neq 0$}

% It would be nice if the author could comment if the AdS black hole solutions exhibited here could be of relevance for investigations in gauge gravity duality.

Let us consider now the effect of a non-vanishing value of $\Lambda$ in the metric function (\ref{metric}). The interest on black holes in AdS spaces is mainly motivated by the AdS/CFT correspondence \cite{AdS/CFT}, which establishes a relation between the thermodynamics of a black hole and the conformal dual field theory lying on the boundary of the AdS space. It should be noted that these nonlinear corrections to the dynamics of the electromagnetic field are expected to modify the physics on the CFT side, which could have an impact, for instance, within applications to holographic superconductors \cite{Holo}. A similar problem to that considered here, for GB theory in vacuum, was studied in \cite{cai1} for AdS spaces and in \cite{cai2} for dS spaces. Moreover, exhaustive analysis of these space-times in the neutral \cite{torii05a} and charged case under Maxwell theory \cite{torii05b} have been also performed. The results obtained in this section extend those for spherical topology horizons $k=1$ to the general class of admissible NED models supporting finite-energy ESS solutions. In our models the equations giving the horizons (\ref{hor}) and the extreme black hole radius (\ref{ebh}) receive $l_{\Lambda}^2$ corrections as

\bea
\frac{l_{\alpha}^2}{2}r_h^{D-5}+r_h^{D-3}-\left(m-\frac{2\kappa^2\varepsilon_{ex}(r_h,q) }{D-2}\right)+\frac{r_h^{D-1}}{l_{\Lambda}^2}&=&0  \\\kappa^2 r_{hextr}^4\left(T_t^t(r_{hextr},q)-\frac{(D-1)(D-2)}{2 l_{\Lambda}^2}\right)&=& \nonumber \\ =\frac{(D-2)(D-3)}{2}r_{hextr}^2 +\frac{(D-2)(D-5)}{4} &l_{\alpha}^2& , \label{ebhl}
\ena
while for the extreme black hole mass these corrections cancel between themselves and Eq.(\ref{ebhm}) remains unmodified. Concerning the metric function $g_{\alpha}(r)$, its behaviour around the center is still dominated by the quantity $\Sigma(q)$ but the large-$r_h$ behaviour is now given by

\be
g_{\alpha}(r)\rightarrow 1+\frac{r^2}{l_{\alpha}^2}\left(1-\left[1-\frac{2l_{\alpha}^2}{l_{\Lambda}^2} \right]^{1/2} \right) , \label{adssol}
\en
and the usual cosmological constant $l_{\Lambda}^2$ is replaced by the effective $l_{eff}^2$ at large $r$, defined as

\be
l_{eff}^2=\frac{l_{\alpha}^2}{1-\left[1-\frac{2l_{\alpha}^2}{l_{\Lambda}^2} \right]^{1/2}}. \label{effcosmo}
\en
However, the signs of $l_{\Lambda}^2$ and $l_{eff}^2$ coincide in all cases, which leads to the usual correspondence between the sign of $l_{\Lambda}^2$ and the asymptotically (A)dS structures. Moreover, the combination of this sign with the one of $l_{\alpha}^2$ leads to four different cases. Let us analyze each case separately, using again the sign of $\Sigma$.

\subsection{Asymptotically AdS solutions ($l_{\Lambda}^2>0$)}

\subsubsection{$l_{\alpha}^2>0$} \label{sub:V}

The term under the square root in (\ref{adssol}) may become negative. Consequently the cosmological constant term must be bounded by below as $l_{\Lambda}^2>2l_{\alpha}^2$ for the metric to be well defined. When $\Sigma(q)>0$ the behaviour of the metric for $D>5$ becomes similar to subcase (\ref{sub:1}), but with an AdS asymptotics $r^2/l_{eff}^2$. Consequently, when $D>5$ this space-time is asymptotically similar to the Schwarzschild-AdS one. For $D=5$ the metric at the center is finite and there are again solutions with a single nondegenerate horizon if $l_{\alpha}^2<2\Sigma$, and two, one (degenerate) or zero horizons otherwise, as in the asymptotically flat case of subsection \ref{sec:IV-I}.

On the other hand, if $\Sigma(q)<0$ the solutions exhibit a BS, becoming similar as those of subcase (\ref{sub:4}) for all $D$, but with an AdS asymptotics as in the previous case.

\subsubsection{$\l_{\alpha}^2<0$} \label{sub:VI}
The term under the square root in (\ref{adssol}) is always positive and thus $l_{\Lambda}^2$ is unbounded. The solutions in this case are similar as those of subcases (\ref{sub:2}), (\ref{sub:3}) and (\ref{sub:5}), again with a AdS asymptotics $r^2/l_{eff}^2$. Note that the typical behaviour of the Reissner-Nordstr\"om-AdS solution is obtained when $D>5, \Sigma(q)>0, l_{\alpha_0}^2<l_{\alpha}^2<0$. It is worth mentioning that in the special case $l_{\Lambda}^2=2l_{\alpha}^2$ the effective cosmological constant of Eq.(\ref{effcosmo}) is precisely $l_{\alpha}^2$ and thus the GB parameter $\alpha$ is identified (times a constant) to the effective cosmological constant of the theory.

\subsection{Asymptotically dS solutions ($l_{\Lambda}^2<0$)}

This case becomes much more involved due to the interplay between $m$, $\varepsilon$, $l_{\alpha}^2$ and $l_{\Lambda}^2$ and, as a consequence, there are several additional black hole structures, already found in the GR-NED system (see e.g. \cite{BI-AdS}). Let us briefly discuss the different possibilities.

\subsubsection{$\l_{\alpha}^2>0$} \label{sub:VII}

Now $l_{\Lambda}^2$ is unbounded. If $\Sigma\geq 0$ and $D>5$ the metric diverges to $-\infty$ both as $r \sim 0$ and at $r \rightarrow \infty$ and we find black holes with a extreme (degenerate) horizon with mass $m=m_{hextr}(q)$, naked singularities ($m<m_{hextr}(q)$), and two-horizons black holes ($m<m_{hextr}$), resembling the usual Schwarzschild-dS behaviour. Similar number of horizons are found when $D=5$ if $l_{\alpha}^2<2\Sigma$, for which the metric at the center  Eq.(\ref{metric}) is negative (but finite). However, when $l_{\alpha}^2>2\Sigma$ the metric at the center (\ref{metric}) is positive (for A0, A1 and A2 fields) and the interplay between $\l_{\alpha}^2$ and $l_{\Lambda}^2$ leads to new kinds of structures formed with three types of horizons (inner, outer and cosmological). Indeed now there exist two kinds of extreme solutions, with masses $m_{hextr}^{(1)}(q)<m_{hextr}^{(2)}(q)$ given by Eq.(\ref{ebhm}). In the former case the degenerate horizon is realized through a joining between the inner and outer horizons (with an additional cosmological horizon) representing the extreme black hole spacetime, while the outer and cosmological ones join in the latter, and there is an additional inner horizon. For masses such that $m>m_{hextr}^{(2)}(q)$ or $m<m_{hextr}^{(1)}(q)$ there is a solution with a single (cosmological) horizon. Finally, for $m_{hextr}^{(1)}(q)<m<m_{hextr}^{(2)}(q)$ we are led to black holes with three non-degenerate horizons. This kind of structures are also found in the GB-Maxwell case \cite{torii05b}.

On the other hand, if $\Sigma<0$, as already stated (see subsection \ref{sec:IV-II}) the metric shows a BS at $r=r_S$, taking a (finite) positive ($>1$) value there for any $D>4$. Consequently the structures in terms of horizons are the same as in the case $D=5$, $l_{\alpha}^2 > 2\Sigma$.

\subsubsection{$l_{\alpha}^2<0$} \label{sub:VIII}

Now $l_{\Lambda}^2$ is bounded as $l_{\Lambda}^2<2l_{\alpha}^2$ and there are three subcases for the metric behaviour. When $\Sigma(q) \geq 0$ the metric at the BS point $g_{\alpha}(r_S)=1+r_S^2/l_{\alpha}^2$ can be either positive or negative, depending on $\vert l_{\alpha}^2 \vert \leq r_S^2$ (its derivative there is always positive): the former case leads to a single-horizon black hole while in the latter there can be black holes with two horizons (event and cosmological), a single one (extremal, the event and the cosmological horizons join) or naked singularities, depending on the mass according to $m\lesseqqgtr m_{hextr}(q)$. Note that, as in the previous cases, $l_{\alpha}^2$ and $r_S^2$ are not independent, being related through equation (\ref{sing}).

When $\Sigma(q)<0$ and $l_{\alpha 0}^2<l_{\alpha}^2<0$, as already seen, the metric diverges around the center when $D>5$ while for $D=5$ it takes a positive finite value $g_{\alpha}(0) \rightarrow 1-\frac{(2l_{\alpha}^2 \Sigma(q))^{1/2}}{l_{\alpha}^2}>1$, but the structure of horizons in both cases is similar as that found in subcase (\ref{sub:VII}) with $\Sigma(q) <0$: extreme solutions $m=m_{hextr}^{(1)}$ (degenerate inner+outer horizon and cosmological horizon) and $m=m_{hextr}^{(2)}$ (inner horizon and degenerate outer+cosmological horizon), black holes with three horizons $m_{hextr}^{(1)}(q)<m<m_{hextr}^{(2)}(q)$ and solutions with a single cosmological horizon for $m<m_{hextr}^{(1)}(q)$ or $m>m_{hextr}^{(2)}(q)$. Finally the case $\Sigma(q)<0,l_{\alpha }^2<l_{\alpha 0}^2<0$ shows two branch singularities and, since the region beyond the outer BS radius is the physically relevant, the structure of horizons is similar as in case $\Sigma<0, l_{\alpha}^2>0$.

\section{Conclusions}

In this paper the families of NED models, constrained by several physical admissibility requirements, and supporting ESS solutions whose energy in flat space is finite, were considered in the framework of Gauss-Bonnet theory. This is the simplest nontrivial extension of General Relativity incorporating higher-order curvature invariants and leading to second-order field equations. The finite character of the energy of the ESS field is established according to the asymptotic and central-field behaviour of the ESS field in absence of gravity. In the latter case two field behaviours compatible with the finite-energy requirement were determined: one divergent at the origin and the other one attaining a finite value there.

With these results we have shown that the associated gravitating structures for these NED models, when coupled to the Einstein-Gauss-Bonnet action, can be qualitatively characterized in terms of the relation between the mass parameter $m$ and the ESS energy $\varepsilon(q)$. These structures fall into two classes according to the Gauss-Bonnet parameter $l_{\alpha}^2$, one defining solutions whose domain of existence is the whole spacetime, and another for which the solutions are not defined everywhere, showing a singularity at a finite radius. In the former case the nature of the ESS field at the center (A0, A1, A2) and the space-time dimension $D$ critically affects the metric behaviour. Indeed, the case $D=5$ was shown to posses a different structure as compared with the remaining cases $D>5$. In addition, when a cosmological constant term is introduced in the system, then the relation between the module and sign of $l_{\alpha}^2$ and $l_{\Lambda}^2$ determines the effective value of the cosmological constant of the corresponding asymptotically (Anti-)de Sitter space. While in the former case the number and kind of gravitating structures remain unmodified, in the latter new structures may appear, including black holes with three non-degenerate horizons, or black holes with a both a degenerated and a nondegenerate horizon. The analysis performed here goes beyond previous results obtained in the literature corresponding to several particular Lagrangian densities \cite{BI-Lovelock}, and extends them to the family of physically admissible gravitating NED models with finite-energy ESS solutions in the context of Gauss-Bonnet theory.

Although we have only analyzed here black holes with the usual spherical topology for the horizon, it is possible to extend these results to the cases of negative constant or zero curvature hypersurfaces. Also a suitable extension of this work would be the analysis of the thermodynamic features of these solutions, concerning also those with the different horizon topologies mentioned above. Indeed, it is well known that while usually entropy of black holes equals one quarter of the horizon radius, this property does not hold, in general, for higher-order curvature theories \cite{Entropy}. Thus, it would be worth studying such features, in a similar way as is done in \cite{cai1,cai2} for (A)dS-GB theory in vacuum and with a Maxwell field. It is expected that, analogously to the geometric study carried out in this paper, thermodynamics of the corresponding solutions will also depend on a few data and, therefore, a similar systematic analysis will also be possible. Such investigations, for non-vanishing cosmological constant term, could be of potential relevance within the framework of gauge/gravity dualities. At this point of our research, and in absence of specific settings on the CFT side, it is difficult to foresee what kind of particular NED theory would be useful on the gravity side. This further supports the interest on methods like those developed here to characterize the generic geometric and thermodynamic behaviour of physically consistent NED theories in asymptotically AdS space-times, which could prove useful in the future to match particular CFT settings.

To conclude, since Gauss-Bonnet and, more generally, Lovelock gravities, are at the crossroad of the metric and Palatini formulations \cite{Borunda}, the study of these higher-order gravity theories and their solutions may reveal useful information on the effective approaches to quantum gravity. A final remark concerns the fact that GB theories coupled to admissible NED models are unable to provide a solution free of curvature singularities everywhere, a reminiscent situation as that of GR, where the central singularity is only avoided (for purely electric fields) through adhoc unphysical choices of the NED Lagrangian density \cite{Regular2}. This motivates further investigations on the avoidance of curvature singularities in other scenarios of modified gravity.

\section*{Acknowledgments}

This work has been supported by the NSFC (Chinese agency) grant Nos. 11305038 and 11450110403, the Shanghai Municipal Education Commission grant for Innovative Programs No. 14ZZ001, the Thousand Young Talents Program, and Fudan University. The author also acknowledges partial support from CNPq (Brazilian agency) grant No. 301137/2014-5 and from the Department of Physics at University of Valencia.

{}

\end{document}